\documentclass[aps,showpacs,twocolumn]{revtex4}
\usepackage{epsfig}
\usepackage{times}
\usepackage[T1]{fontenc}

\newcommand{\beq}{\begin{equation}}
\newcommand{\eeq}{\end{equation}}
\newcommand{\bea}{\begin{eqnarray}}
\newcommand{\eea}{\end{eqnarray}}

\newcommand{\veps}{\varepsilon}

\newcommand{\s}{\sigma}

\newcommand{\goto}{\rightarrow}

\newcommand{\rar}{\rightarrow}

\begin{document}

\title{Emergence of a negative charging energy in a metallic dot
capacitively coupled to a superconducting island}
\author{C. Holmqvist $^{a)b)}$, D. Feinberg$^{a)}$, and A. Zazunov$^{c)}$}
\affiliation{$^{a)}$ Institut NEEL, Centre National de la Recherche
Scientifique and Universit\'e Joseph Fourier, BP 166, 38042 Grenoble,
France}
\affiliation{$^{b)}$Applied Quantum Physics Laboratory, MC2, Chalmers University of Technology, S-412 96
G\"oteborg, Sweden}
\affiliation{$^{c)}$Laboratoire de Physique et Mod\'elisation des
Milieux Condens\'es,
Universit\'e Joseph Fourier and CNRS, BP 166, 38042 Grenoble, France }

\date{\today}
\begin{abstract}
We consider the hybrid setup formed by a metallic dot, capacitively coupled to a superconducting island S connected to a bulk superconductor by a Josephson junction. Charge fluctuations in S act as a dynamical gate and overscreen the electronic repulsion in the metallic dot, producing an attractive interaction between two additional electrons. 
As the offset charge of the metallic dot is increased, the dot charging curve shows positive steps ($+2e$) followed by negative ones ($-e$) signaling the occurrence of a negative
differential capacitance. A proposal for experimental detection is given, and potential applications in nanoelectronics are mentioned.
\end{abstract}


\maketitle

At low temperatures, the electronic transport through point-like metallic nanostructures (quantum dots) is dominated by the electronic Coulomb {\it repulsion} between additional electrons. When a small-capacitance dot is weakly coupled to a normal metallic reservoir, the average number of charges 
in the dot, $n_N$, increases one by one with the gate voltage $V_{gN}$, leading to conductance peaks \cite{cond_peaks}. This Coulomb blockade phenomenon has recently enabled an individual control of charge or spin, for instance in view of quantum information protocols \cite{QI_charge,QI_spin} .  
We address here the possibility of inverting the sign of the charging energy. Indeed, creating a {\it negative} effective charging energy in a normal (non-superconducting) metallic dot would induce attractive correlations, triggering for instance pair tunneling from/to a normal reservoir \cite{vonoppen} or a charge Kondo effect \cite{charge_kondo,Koch}, or giving rise to super-Poissonian shot noise \cite{Koch,Hwang}. Going beyond the single dot case, attractive correlations between electrons in spatially {\it separated} dots could help implementing quantum information protocols involving two-qubit gates, when the qubits are carried by the charge (spin) of the last added electron. Historically, attractive interactions in the solid state  are known as valence-skipping states \cite{valence_skipping}, and negative-U centers \cite{negative_U}. Another possible mechanism for electronic attraction is mediated by optical phonons, binding two electrons as a bipolaron in confined geometries for strong electron-phonon coupling and lattice polarizability \cite{bipolarons1}. Due to the low polarizability and small effective carrier mass, bipolarons are unlikely to form in a clean GaAs/AlGaAs two-dimensional electron gas (2DEG), although they might do so in presence of a few donor impurities \cite{pair_tunneling}, or 
with more polarizable multilayer materials \cite{bipolarons2}.  Molecular junctions are also promising for achieving a negative charging energy \cite{alexandrov,vonoppen}.

In the present Letter, we propose an alternative mechanism, pointing out that the repulsive charging energy in a metallic dot (N) connected to a normal reservoir (Fig. 1) can be turned into an attractive one when N is {\it capacitively} coupled to a superconducting island (S). The latter is connected to a superconducting reservoir by a Josephson junction (JJ) and operates in the Cooper pair box regime, e.g. it fluctuates between two pair number states \cite{bloch_bands,CB}. Here we assume that electron tunneling between S and N is negligible; therefore no proximity effect occurs in the N dot. We instead focus on the charging properties of the N grain as its gate voltage is varied. The S island acts as an effective dynamical gate, whose effects turn out to be nonlinear.  
As the main result of this Letter, the Coulomb charging energy in N can be overscreened by the neighboring pair fluctuations in S, and an effective {\it local attraction} appears between electrons added into N. As a corollary, certain charge states are "skipped" as the N gate voltage is varied. The resulting charging 
curve becomes non-monotonous, displaying positive steps ($+2e$) followed by {\it negative} ones 
($-e$). 
 A related screening effect was proposed by Averin and Bruder for controlling 
 the coupling between two superconducting charge qubits
\cite{averin_bruder}. Notice that if N were
coupled to both drain and source reservoirs, our set-up
would be similar to a Cooper pair box  coupled to a
single-electron transistor (SET). The latter  has been studied in great detail
as a read-out device for a superconducting (charge) qubit embodied in
the S island \cite{qubits_chalmers}. In this case, the capacitive coupling
between N and S must be small, in order to minimize the
decoherence due to backaction of the normal part of the device onto
the superconducting one, whereas in our proposal, the coupling is very strong. 

\begin{figure}[h]
\scalebox{0.38}{\includegraphics{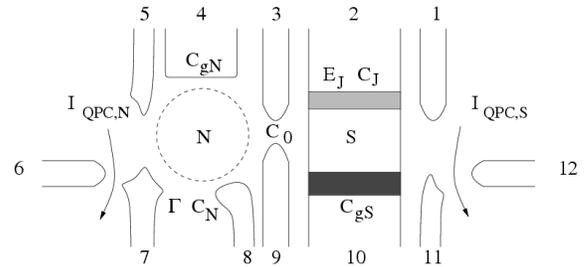}}
\caption{Schematic view of a normal grain (N) coupled with a 
strong capacitive coupling (controlled by 3, 9)
to a Cooper pair box composed of a Josephson junction connecting
superconducting reservoir 2 and island, and gate 10.  Electrons tunnel between N and 
its reservoir (defined by 7, 8). Detection is made by sweeping 
the gate voltage 4 and measuring the island voltages using quantum 
point contacts for both N (5,6,7) and S (1,11,12).}
\label{expt}
\end{figure}

\begin{figure}[h]
  \begin{tabular}{cc}
    \scalebox{0.38}{\includegraphics{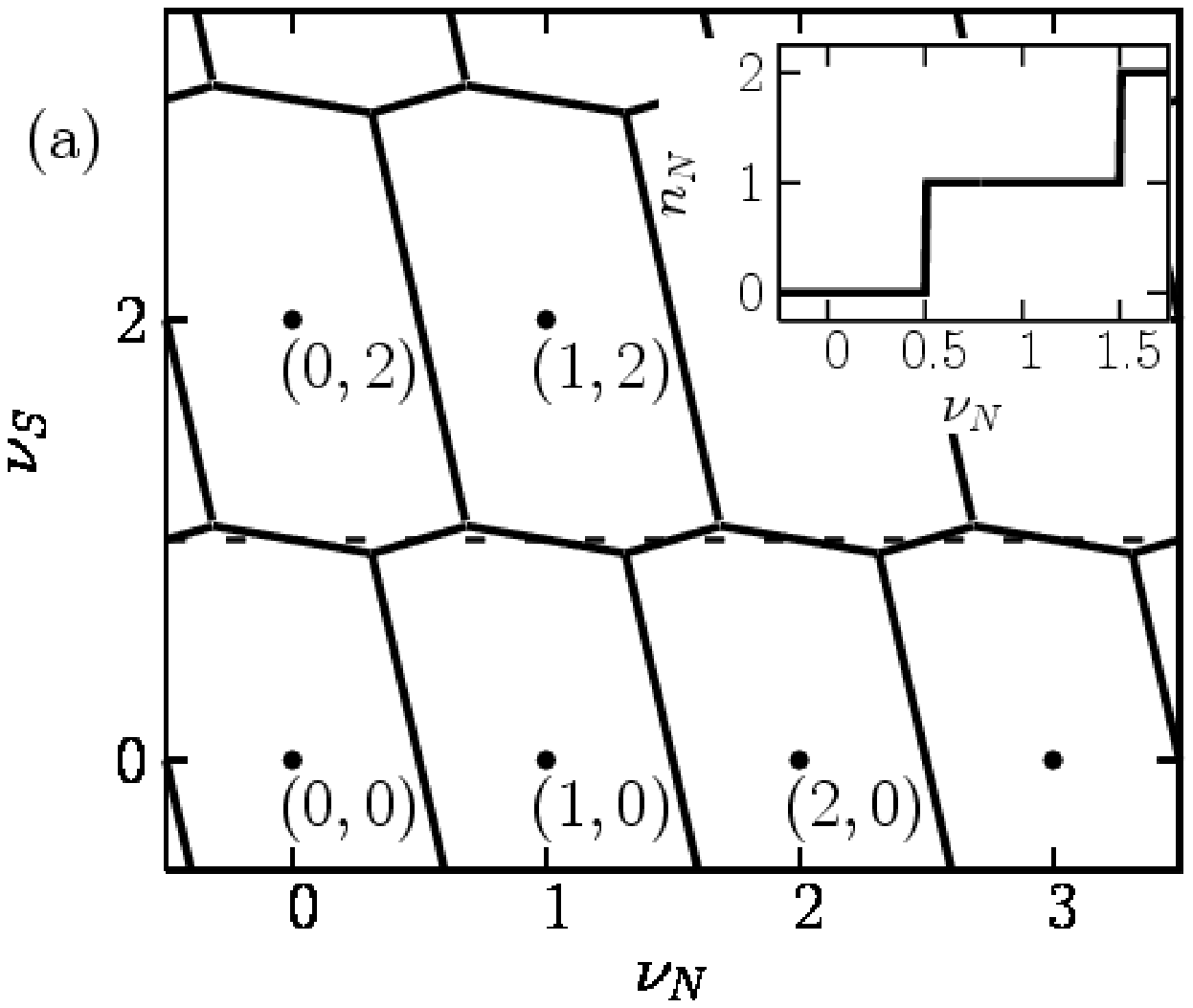}} \\
    \scalebox{0.38}{\includegraphics{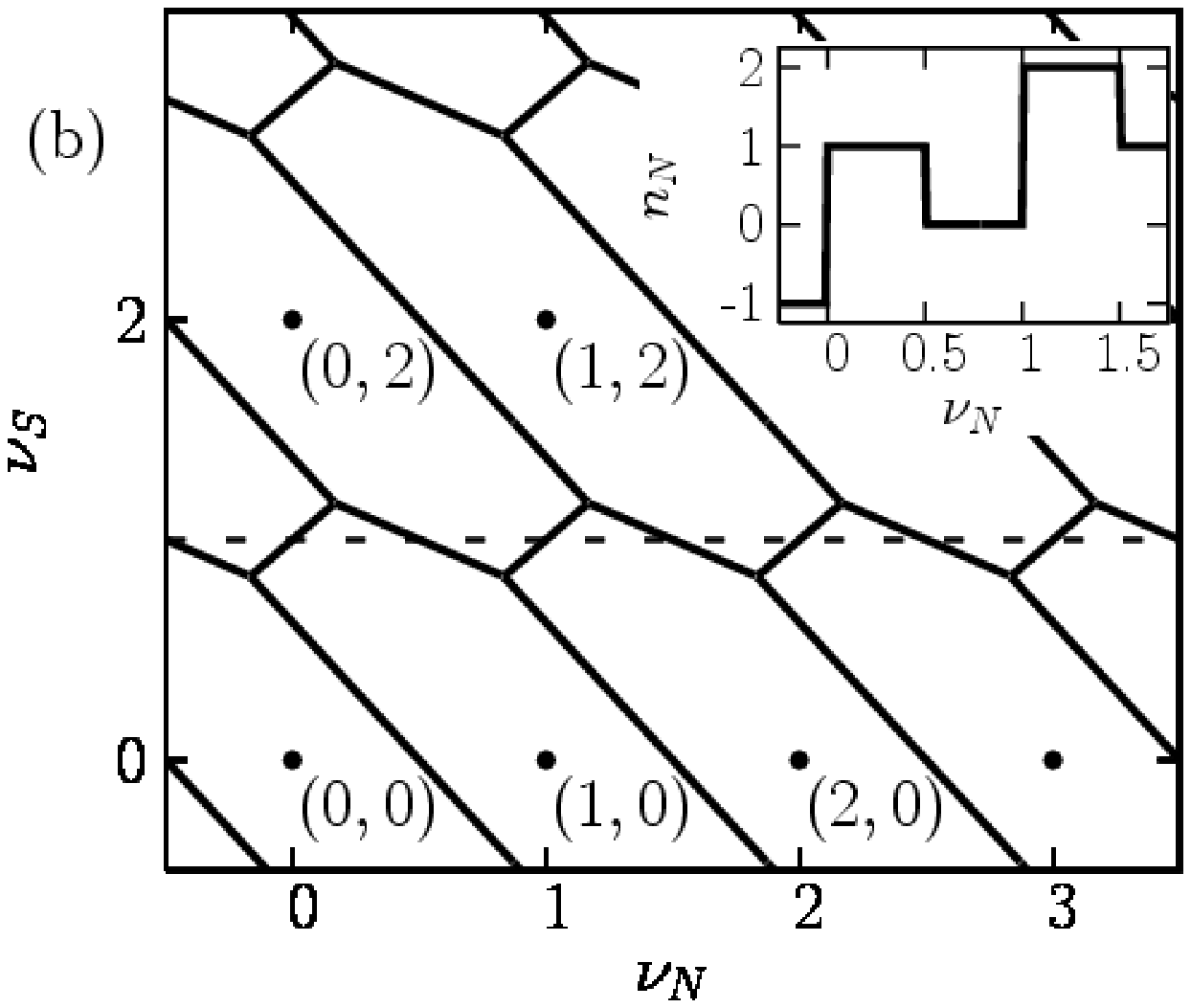}} \\
  \end{tabular}
    
  \caption{Charge stability (or honeycomb) diagram for $b=1$, $r=0.2$ (a) and $r=0.8$ (b). 
The insets show the charging curves for N, taken along the dotted line. In case (b) 
the charging staircase (inset) exhibits charge skipping effects.}
  \label{csd}
\end{figure}

The JJ connecting the S island to the reservoir has a Josephson
energy $E_J$ and capacitance $C_J$, and a gate imposes an
offset $\nu_S=C_{gS}V_{gS}/e$,
with $C_{gS} \ll C_J$. Similarly, the N island is connected to a
normal reservoir by a tunnel junction, with single-electron tunneling
rate $\Gamma$ and capacitance $C_{N}$, and experiences a gate offset
$\nu_N=C_{gN}V_{gN}/e$, with $C_{gN} \ll C_{N}$. Most importantly, the
islands N and S are coupled by a large capacitance $C_0>C_N,C_J$. We
assume the superconducting gap in S to be larger than the charging energy, such that only even
charge number states $n_S$ occur in S. At low temperatures, quasiparticle
tunneling in S can be neglected. Defining $C_{\Sigma S}=C_J+C_0+C_{gS}$ and $C_{\Sigma
N}=C_N+C_0+C_{gN}$, and introducing the parameters 
$b=C_{\Sigma N} / C_{\Sigma S}$ and
$r=C_0 / \sqrt{C_{\Sigma N}C_{\Sigma S}}$, the total charging
energy of the NS system can be written as \cite{DD}

\begin{eqnarray}
\nonumber
E_{C}=E_{CN}[(n_N-\nu_N)^2+b(n_S-\nu_S)^2 
\hspace{1cm} \\
+ \; 2r\sqrt{b}(n_N-\nu_N)(n_S-\nu_S)]
\label{charging}
\end{eqnarray}

\noindent
with $E_{CN}=e^2 / [2C_{\Sigma N}(1-r^2)]$. The asymmetry parameter $b$ and the coupling
parameter $r < 1$ are not independent, and $r < \min{(b, 1/\sqrt{b})}$. 
Eq.~(\ref{charging}) determines the charge stability diagram of the isolated NS system in the
$(\nu_N, \nu_S)$ plane. First, for a value $\nu_S$ imposing an integer
number of pairs in S, say $\nu_S=2$, the charging number $n_N$
increases monotonously with $\nu_N$. Next, consider a case where
$n_S$ fluctuates, for instance $\nu_S=1$. For small $r$, as shown in
Fig. 2a, $n_N$ is again a monotonous
function of $\nu_N$: the sequence of charge states $(n_N, n_S)$
as $\nu_N$ increases reads
$(0,0),(0,2),(1,0),(1,2),(2,0),(2,2), \dots$
(notice the oscillation of $n_S$). The result is very
different if $r$ is large. In fig.~\ref{csd}(b), for
$\nu_S=1$, $n_N$ increases with $\nu_N$ but in a non-monotonous way,
the charge state sequence being
$(1,0),(0,2),(2,0),(1,2),(3,0),(2,2)$,
etc. The corresponding charging staircases are plotted in the insets.

One sees that the transition from $(n_N,2)$ to
$(n_N+2,0)$ at $\nu_N=n_N+1$  ``skips'' the charge state
$n_N+1$ in the grain. This signals a negative effective charging energy in N 
which overcomes the Coulomb
repulsion. After increasing by two units, $n_N$
decreases by one unit, yielding a negative differential
capacitance (NDCA) $C_{diff}=C_{gN} (dn_N/d\nu_N)$ at
half-integer values of $\nu_N$. Strikingly, the total number of steps,  positive or negative,
is doubled with respect to the usual case. Both charge skipping and
NDCA occur above the dotted line indicated in Fig.4 displaying a
$(b,r)$ diagram. 
From the above charging energy, an effective attractive potential $U<0$ can be estimated for $\nu_S=1$ as $U=E_C(0,2)+E_C(2,0)-2E_C(1,2)=e^2(1-2r\sqrt{b})/[C_{\Sigma N}(1-r^2)]$. The necessary condition for the occurrence of a negative charging energy is thus $2r\sqrt{b}>1$.

To further analyze this possibility in an open NS system, let us consider the full Hamiltonian:
 
\begin{eqnarray}
\nonumber
H = E_C + \sum\limits_{k \s}
\veps_{k}\,c^{\dagger}_{kR,\s}\,c_{kR,\s}\,+\,\sum\limits_{q \s}
\veps_{q}\,c^{\dagger}_{qN,\s}\,c_{qN,\sigma}\,\\
+\, [\sum\limits_{k q \s} T_{k,q}\,c^{\dagger}_{kR,\sigma}\,c_{qN,\sigma}  -
\frac{E_J}{2}\,|n_S+2\rangle
\langle n_S|+{\rm H. c.}] ~,
\end{eqnarray}

\noindent
where $k$ ($q$) denotes electron states in the normal reservoir
R (grain N), and 
the Coulomb interaction $E_C$ is given by eq.~(\ref{charging}). The total charge in N is expressed as
$n_N=\sum\limits_{q \s}\,c^{\dagger}_{qN,\sigma}\,c_{qN,\sigma}$. 
Assuming constant densities of
states in N and R, the single-electron
transition rate from R to N is given
by $\Gamma^{(+1)}=[\delta E_C^{(+1)}/e^2R_N]\,[\exp{(\delta
E_C^{(+1)}/k_BT)}-1]^{-1}$ within the golden rule approximation, where $R_N$ is the tunnel resistance.

\begin{figure}[h]
\scalebox{0.38}{\includegraphics{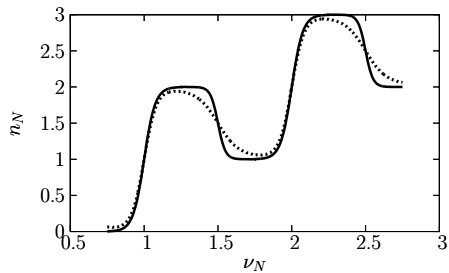}}
\caption{Charging staircase : $r=0.8,\,b=1, \,E_{CS}=E_{CN}, \,k_B T/E_{CN} = 3 \cdot 10^{-2}, \, R_N/R_K = 10$; (full line) $E_J/E_{CN} = 0.5$ ;  
(dotted line) $E_J/E_{CN} = 2$.}
\label{steps_smallEJ}
\end{figure}

Considering first the case of small $E_J \ll E_{CS}=e^2/[2C_{\Sigma S}(1-r^2)]$, 
we perform a T-matrix calculation of the transition rates from
$(0,2)$ to $(2,0)$ (close to $\nu_N=1$) and from
$(2,0)$ to $(1,2)$ (close to $\nu_N=1.5$). For the first
transition, we take into account three possible configuration paths involving
higher-energy states: $(0,2) \rar (1,2) \rar(2,2)
\rar (2,0)$, $(0,2) \rar (1,2) \rar (1,0) \rar
(2,0)$, and $(0,2) \rar (0,0) \rar (1,0) \rar
(2,0)$. For the second transition, only one excited state is
involved: $(2,0) \rar (1,0) \rar (1,2) $ and
$(2,0) \rar (2,2) \rar (1,2)$.
The shape of each step is calculated at finite temperature by
solving the master equation governing the dynamics of the
probabilities $p(0,2)$, $p(2,0)$ for the positive step and
$p(2,0)$, $p(1,2)$ for the negative one. The master
equation reads $\dot{p}(a)=\Gamma^{b\goto
a}p(b)-\Gamma^{a\goto b}p(a)$ with $p(b)=1-p(a)$ for the states
$a,b$ involved in the transition. Here the probabilities of other
states are neglected, which is justified close to $\nu_N=1$ or $\nu_N=1.5$ and 
if the steps are sufficiently narrow.
The resulting steps are shown in Fig.~\ref{steps_smallEJ}.

\begin{figure}[h]
\scalebox{0.38}{\includegraphics{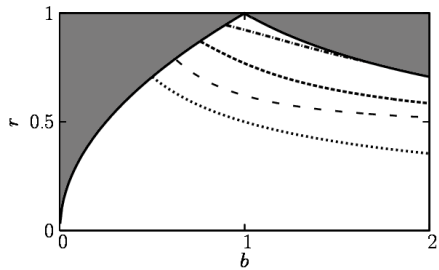}}
\caption{
Phase diagram in the $b,r$ plane. The unphysical grey region is excluded. Charge skipping and NDCA 
occur above the dotted line, from bottom to top : $E_J/E_{CN} = 0$ , and $1, 2, 4$ (adiabatic calculation). All other parameters are the same as in fig. 3.}
\label{steps_largeEJ}
\end{figure}

As a result, a positive step $+2e$
(where the charge number $n_N=1$ is skipped) and a consecutive
negative step $-e$ are stabilized. Notice that contrary to the usual
case where all transitions between $n$ and $n\pm1$ are real and 
obey the same master equation \cite{beenakker}, here the rates are of higher order and the
virtual states involved in one transition become real
states (with first order rate) for the next one. A full treatment of all processes 
is beyond the scope of this Letter.

Let us now turn to the case of a large Josephson energy, 
$E_J > E_{CS}$. The adiabatic
assumption \cite{averin_bruder}, 
setting the phase difference to $\phi$ across the JJ, 
allows to solve the Hamiltonian (2),
neglecting the normal electron tunneling term. 
%fixing the charge number $n_N$. 
The resulting adiabatic Hamiltonian $H_{ad}=E_C -E_J \cos{\phi}$ 
describes a Cooper pair box with an effective gate voltage,
which is an adiabatic function of $n_N$. 
In the tight-binding limit $E_J/E_{CN}\gg b$,
assuming that the junction dynamics is confined to the lowest Bloch band,
one obtains the sum of the N dot charging energy and the adiabatic Bloch band energy :
\bea
\nonumber
H_{ad}\,=\,E_{CN} (1-r^2)(n_N-\nu_N)^2
\hspace{2.5cm} \\
-\,\Delta_0 \cos[\pi (\nu_S\,-\,\frac{r}{\sqrt{b}}(n_N-\nu_N))] ~,
\eea
where the bandwidth is given by \cite{bloch_bands}
\bea
\Delta_0 = 16 \sqrt{\frac{2}{\pi}} \, b \, E_{CN} \left( \frac{E_J}{2 b E_{CN}}
\right)^{3/4}
e^{- \sqrt{8 E_J / b E_{CN}}} ~.
\eea
The second term in $H_{ad}$ represents a nonlinear screening potential acting on the
charge in N. The offset $\nu_S$  controls the phase of the cosine
term, and an appropriate choice (for instance $\nu_S \approx 1$) achieves a 
negative curvature of $H_{ad}$, viewed as an
effective charging energy $E_{CN}^{eff}$ for the gauged charge in N, $n_N-\nu_N$. 
The required condition reads $(\pi^2/2) r^2\Delta_0 /[b (1-r^2)]> 1$,
yielding the lines in Fig. 4. Clearly, a large $E_J$ puts a stronger constraint on the coupling capacitance $C_0$, and requires larger values of $r$ than for small $E_J$. The shape of the charge skipping and negative steps 
is then calculated, using
a master equation based on charge states
$n_N=0,2$ or $n_N=2,1$, respectively. The adiabatic transition
rates are given by $\Gamma_{ad}=[\delta
E_{CN}^{eff}/e^2R_N]\,[\exp{(\delta E_{CN}^{eff}/k_BT)}-1]^{-1}$.
The corresponding steps (Fig. 3) are less pronounced than in the small $E_J$ case.

Searching the optimum regime must account for the fact that for too large $r$ values, the system behaves like one single island and its energy no longer depends on the location of the charge. The Coulomb blockade requires temperatures much smaller than the energy difference between two charge states.  An optimal $r$ is close to $0.75$ (with $b=1$) for small Josephson energies, and the requirement for Coulomb blockade reads $k_B T < E_{CN}/4$. The value $r = 0.8$ was used 
in the steps calculations. 
A temperature of $T \sim 30$ mK and a typical charging energy of $E_{CN} \sim 0.1$ meV were used, yielding $C_N = C_J \sim 2$ fF and $C_0 = 4 C_N = 8$ fF. The tunnel resistance $R_N$ is chosen as $R_N / R_K = 10$ with $R_K = h/e^2 \approx 25.8$ k$\Omega$. This yields a 
bare single-electron tunneling rate of $\Gamma \sim 10^{9}$ s$^{-1}$. 
In presence of the S island, the effective tunneling rates are $10^{7}$ s$^{-1}$ ($n_N$ decreasing from $2$ to $1$) and $5 \cdot 10^{3}$ s$^{-1}$ ($n_N$ increasing from $0$ to $2$), respectively. With the above parameters, the maximum attraction $|U|$ is of the order of $50 \mu$eV. 

Notice that such a negative charging energy cannot render the dot superconducting, because it concerns the energy required to add one or two electrons on the dot, rather than 
a true attractive potential felt by all electrons near the Fermi level. The effective "negative-U" potential manifests itself only when the dot is weakly coupled to a reservoir such that its charge can fluctuate. This is similar to a single "$-U$" center weakly hybridized with a normal bulk metal \cite{negative_U}. Here, pair fluctuations with the reservoir are assumed to be incoherent. On the contrary, a very small tunneling term $T_{NS}$ between N and S  would allow for a true phase coherence between states
$n_N$, $n_N+2$, thus a proximity effect in a quite unusual regime, where $T_{NS} < |U|$.

Let us briefly discuss the issue of phase coherence in the Cooper pair box. 
Charge skipping only requires pair tunneling between the superconducting reservoir and the S island in
order to screen the repulsive interaction in the normal grain. 
No phase coherence is needed, as shown by our calculation in the small 
$E_J$ case. Moreover, even in the large $E_J$ case, charge fluctuations in N should
strongly react back upon S and reduce the phase coherence. A full treatment 
goes beyond the adiabatic approximation \cite{averin_bruder}. 
One can anticipate that  large fluctuations in the phase $\phi$
renormalize $E_J$ downwards, making the small-$E_J$ case generic. 

We now propose a scheme for detecting the non-monotonous charging of the N
grain. SETs, or point contacts \cite{field} provide very sensitive
detection of the local change in the electrostatic potential. 
In double-dot setups with weak mutual coupling,
the potential variations in each dot can be measured by a different
neighboring point contact \cite{DDN2}. In the present case,
placing a point contact close to N does not measure $\delta n_N$, but
instead $\delta V_N=(C^{-1})_{NN}(e \delta
n_N)\,+\,(C^{-1})_{NS}(e \delta n_S)
=e\,[\delta n_N\,+\,r\sqrt{b}\,\delta
n_S]/[C_{\Sigma N}(1-r^2)]$. If $2r\sqrt{b}>1$, doubling of the number of steps
can be detected, but not the non-monotonous charging curve. To access
the latter, it is suitable to measure $\delta V_S=e\,[r\sqrt{b}\,\delta n_N\,+\,b\,\delta n_S]/[C_{\Sigma
N}(1-r^2)]$ as well, with
a second point contact close to S, and reconstruct
$\delta n_N=C_{\Sigma N}\,[\delta
V_N-r\delta V_S/\sqrt{b}]/e$. The parameters $C_{\Sigma N},r,b$
can easily be measured from the stability diagram obtained in the normal state in the presence of a very weak tunneling between N and S \cite{DD}. Notice that the 
tunneling rates calculated above are much reduced compared 
to the bare single-electron rate $\Gamma$. 
Therefore the use of point contacts permits, non only a time-averaged \cite{DDN2}, but even a time-resolved and directional \cite{fujisawa} detection of the charge variations in N and S. 
On the other hand, cross-correlation shot noise measurements, as in ref.~\cite{cross_corr}, would require higher currents. In practice, a possible setup inspired by ref.~\cite{DDN2}
is proposed in Fig.~\ref{expt}. It involves a
superconducting strip with a Cooper pair box, coupled laterally to an InGaAs/AlGaAs
2DEG, suitably tuning the Schottky barrier present at the interface between the superconductor and the 2DEG. Notice that the geometry of Fig. 1 could be modified, including drain and source such as to allow transport through N \cite{vonoppen}.

In conclusion, we have proposed a mechanism inducing a controllable negative charging energy, thus attractive correlations, in one or several metallic dots. We believe that such an effect would be useful in view of more complex nanoelectronics devices. 
The authors are grateful to T. Martin, M. Fogelstr\"om, and G. Johansson for useful discussions, and S. Andergassen for careful reading of the manuscript. D. F. and A. Z. were partially supported by the contract AC NANO NR0114, and C. H. was supported by the Swedish Research Council (VR) under grant 621-2006-3072.

\end{document}